\documentstyle[12pt]{article}
\begin{document}
\setlength{\baselineskip}{24pt}
\title{Quadrupole Contribution to Two Neutron Removal in Heavy Ion
Collisions}
\author{C.J. Benesh, \\
T-5, MS B283 Los Alamos National Laboratory \\
Los Alamos, NM 87545 }
\maketitle

\begin{abstract}
	In this report, electric quadrupole corrections to the
two neutron removal cross section measured in heavy ion collisions
are estimated for $^{197}$Au and $^{59}$Co
targets. The quadrupole process is assumed to proceed primarily through
excitation of the giant isovector quadrupole resonance, which then
decays by neutron emission. For $^{59}$Co, the contribution from E2
radiation is found to be small, while for $^{197}$Au we find the quadrupole
contribution resolves the discrepancy between experiment and the simple
predictions of the Weissacker-Williams virtual photon method.

\end{abstract}

	The combination of large charges and enhancement of the
radiation
pulse at high energies by Lorentz contraction make Electromagnetic
Dissociation(ED) by heavy ions an attractive method for measuring low
energy
nuclear cross sections of astrophysical interest\cite{1}. At the same
time, the
very large cross sections, on the order of barns, attainable in ED
will
contribute substantially to the hadronic background  at RHIC\cite{2}.
A thorough understanding of the ED process is necessary to extract
meaningful
results from these experiments. It is not clear, however, that such
an
understanding exists at present. Indeed, it has been found in target
fragmentation experiments that the single and double neutron removal
cross sections do not agree with the simple Weissacker-Williams(WW) theory
of the ED reaction\cite{2,3,4,5,6}. For the most part, the discrepancies in
the single neutron removal process have been resolved by careful
consideration of the
choice of critical impact parameter\cite{7}, inclusion of quadrupole
excitations in the EM cross section\cite{7}, and the effect of the
Rutherford bending of the projectile's trajectory\cite{8}. Recently,
these considerations have begun to be applied to the two neutron
removal process as well\cite{9}. In this report,
we examine the role played by quadrupole excitations, in particular,
the Isovector Giant Quadrupole Resonance(IVGQR), in the two neutron
removal process.

	To date, the most thorough study of the systematics of the
two neutron removal process have been carried out by the Iowa State
Group\cite{2}.
For a variety of projectiles, they have measured the two neutron
removal
cross sections on $^{197}$Au and $^{59}$Co targets(see table). Using
factorization\cite{10} to
estimate the piece of the cross section due to nuclear processes,
they
extract the electromagnetic contribution and find sizeable
discrepancies
from the predictions of the simple WW calculation for the Au target.

In reference 9, Norbury
advances the idea that the discrepancies may be understood if one
assumes
that the uncertainties in the experimental measurements have been
underestimated. As evidence for this, he assumes that the ratio of
the experimental cross sections for one and two neutron removal
should
be independent of the projectile, and then estimates a revised
``experimental'' cross section for $^{197}$Au targets using the
$^{56}$Fe data point, which agrees with the simple WW calculation, to
normalize the
remaining data. The revised cross sections agree to within a few
millibarns with the
naive WW prediction. Underlying this procedure is the assumption that
the
one and two neutron removal cross sections are both dominated by the
naive
WW cross sections, which do have an approximately constant ratio. For
single
neutron removal, this assumption is reasonably valid, as the
corrections
due to quadrupole excitations are on the order of ten per cent for
the projectiles under consideration\cite{7}. This is most likely not
the case for two neutron removal. Unlike the single neutron case, the two
neutron threshold(15 MeV for $^{197}$Au)
lies above the energy of the Giant Dipole Resonance(GDR)(13.8 MeV for
$^{197}$Au), so that
contribution of the GDR is considerably smaller than it is for
single neutron removal. On the other hand, the isovector giant
quadrupole
resonance lies above the two neutron threshold(23 MeV for Au)
and as a result decays primarily into two neutrons. In addition,
for the lowest energy projectiles, where the discrepancy is
largest, the WW quadrupole flux will be enhanced relative
to the dipole flux. Thus, a  reasonable expectation is that the
quadrupole contribution will be a significantly larger percentage of the two
neutron cross section than for single neutron removal.

	In order to estimate this contribution, one needs a model to
separate the quadrupole part of the photo-cross section from the dipole piece.
The quadrupole cross section is assumed to dominated by the IVGQR, with a
correction factor for the two neutron threshold,
\begin{equation}
\sigma_{E2}(E)={ f\sigma_{EWSR}E_{IVGQR}^2\over
(1+(E^2-E_{IVGQR}^2)^2/E^2\Gamma^2)}
{(E-E_{th})^2\over (\Lambda^2+(E-E_{th})^2)},
\end{equation}
where $f$ is product of the fractional saturation of the energy
weighted sum rule and the branching ration for two neutron decay,
$E_{IVGQR}$ is the resonance energy, $\Gamma$ the resonance width,
$E_{th}$ the
two neutron removal threshhold, $\sigma_{EWSR}$ is the energy
weighted sum rule\cite{9},
and $\Lambda$ a parameter which determines how rapidly the
two neutron channel opens. From reference 11 and statistical model
calculations\cite{12},  $f\approx .8$, $E_{IVGQR}=23$ MeV, and
$\Gamma =7$ MeV for $^{197}$Au. For $^{59}$Co, there is no data
on the IVGQR available, so
the parameters for $^{58}$Ni are used( $f\approx .1$, $E_{IVGQR}=29$
MeV, $\Gamma =9$MeV) The parameter $\Lambda$ is determined from
photonuclear data to be roughly 2 MeV for both Co and Au targets.

	The ED cross section is determined by folding the equivalent
photon fluxes\cite{1} over the cross sections. Assuming that only
the first two electric multipoles are important,
\begin{eqnarray}
\sigma_{ED}&=&\int_{E_{th}} dE\, \sigma_{E1}(E)\, n_{E1}(E) +\int_{E_{th}}
dE\,\sigma_{E2}(E)\, n_{E2}(E)\nonumber\\
  &\approx& \int_{E_{th}} dE\, \sigma_{photo}(E)\, n_{E1}(E)
+\int_{E_{th}} dE\,
\sigma_{E2}(E)\, (n_{E2}(E)-n_{E1}(E)),\nonumber\\
&  &
\end{eqnarray}
where $\sigma_{photo}(E)$ is the experimental photoneutron cross
section, and $n_{En}$ is the appropriate photon flux for the nth multipole
from reference 1. The resulting cross sections are shown, along with the
experimental results and the E2 correction, in the table below.

For Co, the additional contribution from the quadrupole
flux is mitigated by the lack of IVGQR strength, and the already good
agreement between theory and experiment remains undisturbed.
For the Au target, the ED cross section is increased by roughly a third
for all projectiles and agreement with the experimental results is
improved
for all projectiles except Fe, which, as noted in reference 9, lies
suspiciously low relative to the other data. The remaining
discrepancies
are for cases were the measured ED cross section is small, and
consequently
more sensitive to possible systematic errors in measurement and/or
extraction
of the nuclear contribution.  Thus, one may conclude that
the discrepancies observed in the two neutron ED cross sections may
be well described by the WW method if the E2 strength present in the
IVGQR resonance is properly accounted for.

\centerline{\bf {Acknowledgements}}
	This work was supported in part by the U.S. Department of
Energy, Division of High Energy and Nuclear Physics, ER-23.

\vfill\eject
\begin{tabbing}
xxxxxxxxxxxx\=xxxxxxxx\=xxxxxxxx\=xxxxxxxxxx\=xxxxxxx\=xxxxxxxx\=xxxx
xxxxxxxx\= \kill
{\underline{Projectile}} \> {\underline{Target}} \>
{\underline{Energy}} \> {\underline{$\sigma^{expt}_{EM}$}} \>
{\underline{$\sigma^{ww}_{EM}$}} \> {\underline{$\delta
\sigma^{ww}_{E2}$}} \> {\underline{$\sigma^{WW}_{EM}$ + $\delta
\sigma_{E2}$}} \\
\hspace*{0.15in} $^{12}C$ \> \hspace*{0.05in} $^{59}$Co \>
\hspace*{0.05in} 2.1 \> \hspace*{0.01in} 6$\pm$4 \> \hspace*{0.03in}
1.1 \> \hspace*{0.03in} 0.05 \> \hspace*{0.30in} 1.2 \\
\hspace*{0.15in} $^{20}Ne$ \> \hspace*{0.05in} $^{59}$Co \>
\hspace*{0.05in} 2.1 \> \hspace*{0.01in} 3$\pm$5 \> \hspace*{0.03in}
2.9 \> \hspace*{0.03in} 0.1 \> \hspace*{0.30in} 3.0 \\
\hspace*{0.15in} $^{56}Fe$ \> \hspace*{0.05in} $^{59}$Co \>
\hspace*{0.05in} 1.7 \> \hspace*{0.01in} 13$\pm$6 \> \hspace*{0.03in}
14 \> \hspace*{0.03in} 0.6 \> \hspace*{0.30in} 14.6 \\
\hspace*{0.15in} $^{139}La$ \> \hspace*{0.05in} $^{59}$Co \>
\hspace*{0.05in} 1.26 \> \hspace*{0.01in} 32$\pm$16 \>
\hspace*{0.03in} 44 \> \hspace*{0.03in} 2 \> \hspace*{0.30in} 46 \\
\hspace*{0.15in} $^{238}U$ \> \hspace*{0.05in} $^{59}$Co \>
\hspace*{0.05in} 0.96 \> \hspace*{0.01in} 80$\pm$24 \>
\hspace*{0.03in} 65 \> \hspace*{0.03in} 4 \> \hspace*{0.30in} 69 \\
\> \\
\hspace*{0.15in} $^{12}C$ \> \hspace*{0.05in} $^{197}Au$ \>
\hspace*{0.05in} 2.1 \> \hspace*{0.01in} 9$\pm$17 \> \hspace*{0.03in}
5 \> \hspace*{0.03in} 2 \> \hspace*{0.30in} 7 \\
\hspace*{0.15in} $^{20}Ne$ \> \hspace*{0.05in} $^{197}Au$ \>
\hspace*{0.05in} 2.1 \> \hspace*{0.01in} 49$\pm$15 \>
\hspace*{0.03in} 14 \> \hspace*{0.03in} 5 \> \hspace*{0.30in} 19 \\
\hspace*{0.15in} $^{40}Ar$ \> \hspace*{0.05in} $^{197}Au$ \>
\hspace*{0.05in} 1.8 \> \hspace*{0.01in} 76$\pm$18 \>
\hspace*{0.03in} 38 \> \hspace*{0.03in} 13 \> \hspace*{0.30in} 51 \\
\hspace*{0.15in} $^{56}Fe$ \> \hspace*{0.05in} $^{197}Au$ \>
\hspace*{0.05in} 1.7 \> \hspace*{0.01in} 73$\pm$13 \>
\hspace*{0.03in} 73 \> \hspace*{0.03in} 25 \> \hspace*{0.30in} 98 \\
\hspace*{0.15in} $^{139}La$ \> \hspace*{0.05in} $^{197}Au$ \>
\hspace*{0.05in} 1.26 \> \hspace*{0.01in} 335$\pm$49 \>
\hspace*{0.03in} 238 \> \hspace*{0.03in} 86 \> \hspace*{0.30in} 324
\\
\hspace*{0.15in} $^{238}U$ \> \hspace*{0.05in} $^{197}Au$ \>
\hspace*{0.05in} 0.96 \> \hspace*{0.01in} 470$\pm$110 \>
\hspace*{0.03in} 430 \> \hspace*{0.03in} 173 \> \hspace*{0.30in} 603
\end{tabbing}
Two neutron removal cross sections for $^{59}Co$ and $^{197}Au$
targets $\sigma^{expt}_{EM}$ from reference 11, $^{238}U$ from
reference 13.  $\delta \sigma^{ww}_{E2}$ is the correction to the
naive Weissacker-William cross section due to the isovector E2
resonance.

\end{document}